\documentstyle[editedvolume,psfig]{crckapb10} 

\begin{opening}
\title{Using new submillimetre surveys to identify the\protect\\
evolutionary status of high-z galaxies}

\author{David Hughes \& James Dunlop}
\institute{Institute for Astronomy, University of Edinburgh,\\
           Royal Observatory, Edinburgh, EH9 3HJ, U.K.}

\end{opening}

\runningtitle{submm surveys \& the 
evolutionary status of high-z galaxies}

\begin{document}

\section{Why submillimetre surveys?}

In this paper we briefly describe a `key' survey at submm 
wavelengths which we are currently 
conducting to address some of the most important questions in cosmology - 
how, at what epoch and over what period of time did massive galaxies
form at high-redshift?

The primary motivation for undertaking surveys of high-z galaxies at 
submm wavelengths is the expectation,
which has still to be proved, that the most massive galaxies (M$_{baryonic}
\ge 5 \times 10^{11}$M$_{\odot}$) form the
majority of their stars in a relatively short ($< 1 Gyr$), but
extremely luminous phase at rest-frame FIR wavelengths and hence,
at redshifts $z > 3$, at wavelengths in the atmospheric
windows between $350\mu - 1300\mu$ which are accessible from the ground.    

Throughout this paper we are 
referring to the {\em rest-frame} FIR emission when discussing the
observed submm spectral energy distribution, and that all
physical quantities are calculated assuming H$_{0} =
50$kms$^{-1}$Mpc$^{-1}$
and $q_{0}=0.5$.

\subsection{Evolutionary status of high-z galaxies}

\begin{figure}
\vspace{8.6cm}
\includegraphics{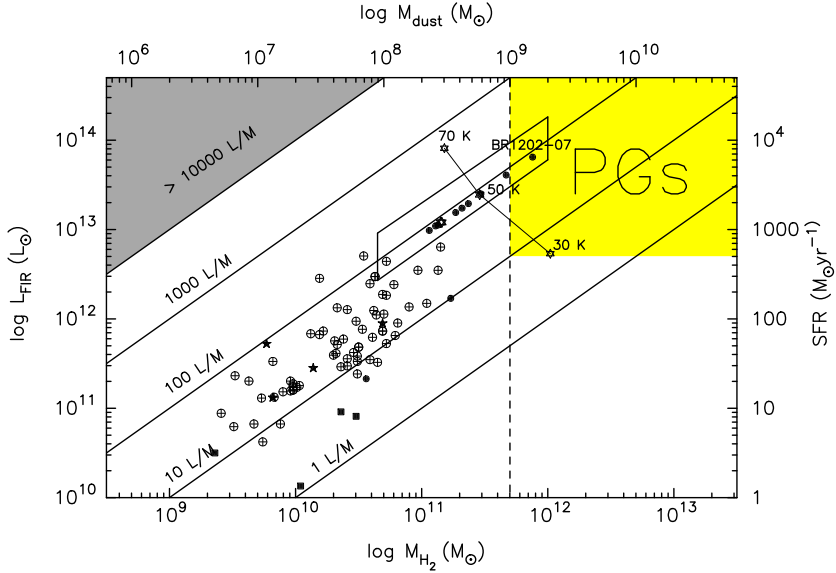}
\caption{\label{fig}
The physical properties of high-z radio galaxies (open stars) and radio-quiet
quasars (filled circles), which lie within the parallelogram, have been 
derived from submm or mm continuum detections (assuming $T = 50$\,K, 
M$_{H_{2}}$/M$_{d} = 500$) and are compared to those of starburst
galaxies (filled stars), ULIRGs (crossed circles) and elliptical
galaxies (filled squares) in the local 
universe. The diagonal lines indicate constant L$_{FIR}$/M$_{H_{2}}$. 
The vertical dashed-line shows the gas mass boundary, to the right of
which is a region of the parameter space marked "PGs" 
where one can expect to find the projenitors of the most massive, 
$> 5 \times 10^{11}$M$_{\odot}$, elliptical galaxies.
This figure is taken from Hughes, Dunlop \& Rawlings (1997).}
\end{figure}

Submm and mm 
continuum observations offer a unique opportunity to determine
some fundamental properties that provide a
measure of the evolutionary status of high-z galaxies,
but only after making the following assumptions:\\

\noindent 
(a) that emission at submm wavelengths is due to 
{\em optically-thin} thermal
re-radiation from dust grains at temperatures of $T_{rest} \simeq 30 - 
70$\,K, and is neither the optically-thin emission of a lower-frequency
radio synchrotron component ({\it e.g} B20902+34 Downes {\it et al.}
1996, Yun \& Scoville - {\em priv. comm.}), 
or the optically-thick self-absorbed emission from a
higher-frequency mid-IR synchrotron component ({\it e.g.} 
de Kool \& Begelman 1989, Schlickeiser {\it et al.} 1991);\\

\noindent 
(b) that the thermal emission from dust grains at an observed
temperature $T_{obs}$ is due to a high-z target, 
where $T_{obs} = T_{rest}/(1+z)$, and is not confused with 
colder foreground galactic cirrus at 15 - 25\,K;\\
   
\noindent 
(c) that within the high-z galaxy, 
the grains are heated by young, massive stars in active
starforming regions, and not by emission from the
associated AGN.\\  

Given the above, it is possible to use the measured submm flux 
densities from high-z galaxies 
to estimate their rest-frame FIR luminosities, which are
proportional to the starformation rate of massive stars at some early
epoch. Additionally the remaining molecular gas mass available
for future star-formation can be determined from a measure of 
the total dust mass, and compared to the expected 
baryonic mass in a present-day counterpart.
For example observational evidence suggests that the most luminous  host 
galaxies ($L \ge 5 L_{\ast}$) 
of low-z quasars and radio galaxies have masses 
$\ge 5 \times 10^{11}$\,M$_{\odot}$ (Taylor {\it et al.} 1996), 
and thus we can describe the
progenitors of similar galaxies forming around luminous high-z AGN 
as {\em prim\ae val} 
if they can be shown to contain a high gas content, comparable to the
stellar mass in low-z quasars and radio galaxies, and 
to be undergoing an intense burst of starformation.
The uncertainties in the dust temperature, 
grain models, gas/dust ratio and their
consequences in the determination of the FIR luminosities, SFRs, and 
gas masses in high-z galaxies have been described in detail elsewhere
by Hughes (1996) and Hughes, Dunlop \& Rawlings (1997). 

A summary of the physical properties of the few
high-z radio galaxies and RQQs detected at 
submm 
and/or mm wavelengths is presented in fig.\,1. These properties 
are compared to those of the most luminous and active galaxies in
the local universe. 
A conservative conclusion is that it is difficult, 
except possibly in the case of BR1202$-$0725 (although it may be lensed)
to describe any of the high-z galaxies detected 
at submm and mm wavelengths as genuinely {\em
prim\ae val}, since they lie outside the region of parameter 
space where, according to our definition of M$_{H_{2}} > 5 \times
10^{11}$\,M$_{\odot}$ and a SFR\,$> 500$\,M$_{\odot}$yr$^{-1}$, young
galaxies should lie.  
Varying the assumed dust temperature through a reasonable range 
($30K \rightarrow 70\,K$), as illustrated by the representative
locus in fig.\,1,
struggles to change this basic conclusion, but it does demonstrate the
need to constrain the dust temperature which can only be achieved by
making submm surveys at two or more wavelengths that
straddle the rest-frame FIR peak at the highest
redshifts (see \S 1.2). 

\subsection{Technical Feasibility of Future Submm Surveys}

\begin{figure}
\vspace{8.7cm}
\includegraphics{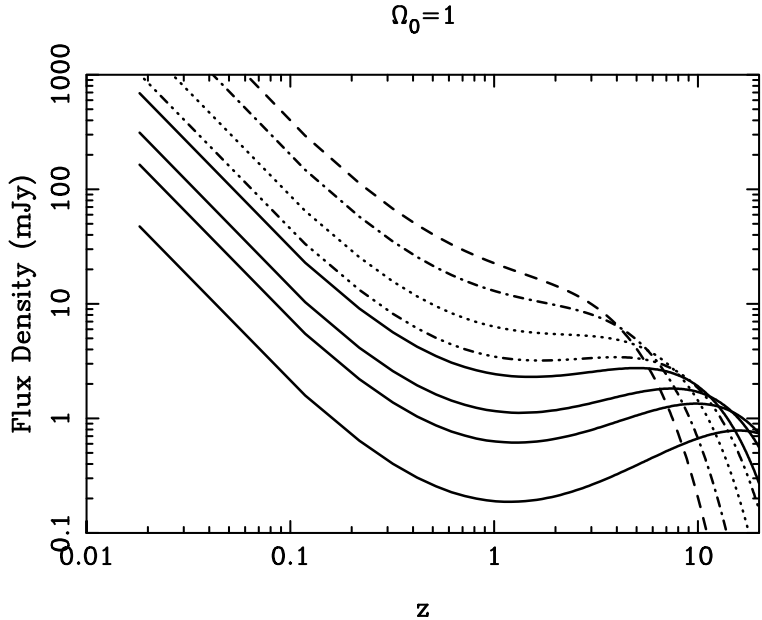}
\caption{\label{fig} 
Each curve
represents the dependence of flux density with redshift for a typical
starburst galaxy spectrum at wavelengths of 350$\mu m$,
450$\mu m$, 600$\mu m$, 750$\mu m$, 850$\mu m$, 1100$\mu m$, 1300$\mu m$
and 2000$\mu m$, from top to bottom respectively. 
The curves are normalised to the flux density of Arp200 at z=0.018 with
a FIR luminosity $L_{FIR} \simeq 2 \times 10^{12} L_{\odot}$ and a 
SFR $ \sim 100$M$_{\odot}$yr$^{-1}$. 
}
\end{figure}

It's all very well describing the importance of submm
observations of high-z sources, but do the sensitivities of 
current and future instruments
and telescopes make the detections of large numbers of high-z galaxies a
realistic possibility? Such concerns are well justified given the
extreme diffculty found in detecting only a handful of high-z AGN
at submm wavelengths without the assistence of gravitational
lensing (Dunlop {\it et al.} 1994, 
Ivison 1995, Isaak {\it et al.} 1994, Chini \& Kr\"{u}gel 1994, 
McMahon {\it et al.} 1994, Cimatti \& Freudling 1995,
Hughes, Dunlop \& Rawlings 1997, 
Omont {\it et al.} 1996).

In fig.\,2 we illustrate the dependence of flux density on redshift for a
typical starburst galaxy SED in the major submm and mm atmospheric
windows. The particular example shown is that of Arp 220 (log L$_{FIR} \sim 
12.3$\, L$_{\odot}$, SFR $\sim 100$\,M$_{\odot}$yr$^{-1}$), 
although the flux densities can be scaled for a starburst galaxy
with  any arbitrary luminosity or SFR at any redshift. The conclusion
that can 
be drawn from fig.\,2 is that submm and mm instrumentation must achieve
NEFD's of $<30$mJy\,Hz$^{-1/2}$, $<80$mJy\,Hz$^{-1/2}$, 
$<250$mJy\,Hz$^{-1/2}$ at 1.3\,mm, 850$\mu m$ and 450$\mu m$
respectively if future studies are to detect, with reasonable
integration times,  high-z galaxies undergoing 
starformation at rates of $>100$\,M$_{\odot}$yr$^{-1}$. 
Such high and sustained SFRs are necessary if a massive galaxy, $\gg
10^{11}$M\,$_{\odot}$, is to be {\em built} in $< 1$Gyr - a timescale 
which, at high-z, 
corresponds to a redshift interval $\Delta z \sim 4.5 - 2.5$.
These necessary 
instrumental sensitivities can be compared with those in table\,1
for all the major existing (and proposed) submm telescopes. It is
immediately clear that only the JCMT at 850$\mu m$ and IRAM at $1.3$mm
currently provide sufficient sensitivity to carry out extensive
cosmological studies, and that further advances must await the future
developments of the SMA, the South Pole 10-m project, FIRST and the
ambitious large mm arrays ({\it e.g.} MMA and LMSA).

\begin{table}[htb]
\begin{center}
\caption{Continuum sensitivities of receivers on
submm/mm telescopes. The SMA(6) NEFDs assume
6 antennae in the array; SMA(8) includes two additional 
Taiwanese antennae, and in parentheses the use of dual polarization 
heterodyne receivers at
345GHz. Numbers in italics are predicted sensitivities.}
\begin{tabular}{lrccc}\hline
 \multicolumn{2}{c}{} & \multicolumn{3}{c}{NEFD (mJy Hz$^{-1/2}$)} \\                              
telescope & area (m$^{2}$) & 1350$\mu m$ & 850$\mu m$ & 450$\mu m$ \\ \hline
JCMT    & 177 \hspace{3mm} & 60  &   80 & 700 \\
CSO     &  79 \hspace{3mm} &      &      & 2000 \\
IRAM    & 707 \hspace{3mm} &  60  &      &      \\
SEST    & 177 \hspace{3mm} & 200  &      &      \\ 
SOFIA   &   5 \hspace{3mm} &      &  {\it 200} & {\it 200} \\ 
SMA(6)  & 170 \hspace{3mm} & {\it 102}  &  {\it 238}  & {\it 2040}  \\
SMA(8)+JCMT+CSO & 481 \hspace{3mm}   & {\it 36}   &  {\it 84 (59)} & {\it 719} \\ 
South Pole 10-m & 79 \hspace{3mm}  &     & {\it 50}  & {\it 60} \\
MMA     & 2010 \hspace{3mm} & {\it 8}  & {\it 25}    & {\it 150} \\
FIRST   &  7  \hspace{3mm}  &    & {\it 54} & {\it 54} \\ \hline
\end{tabular}
\end{center}
\end{table}

In addition to the gains in sensitivity, the availability of new 
instrumentation on ground-based telescopes ({\it e.g.} JCMT, IRAM)
and satellites (ISO) gives rise to a combination of 
improved resolution, imaging capability and greater
wavelength coverage in the FIR, submm and mm. Therefore it should 
soon be possible for the first time to properly 
test the validity of the assumptions 
outlined in \S 1.1, particularly quantifying the level of cirrus 
confusion on scales of $\sim 10''$, constraining the dust temperature, and 
discriminating between the competing thermal and
non-thermal emission mechanisms in the FIR 
by measuring directly, and with high
photometric precision, the rest-frame submm
spectral indices of high-z galaxies (Chini {\it et al.} 1989, 
Hughes {\it et al.} 1993). 
Finally, since all massive galaxies at high-z have been originally
identified by detecting a luminous AGN 
then the effect of the AGN continuum
emission, through primary or secondary processes, on the FIR luminosity
and on the overall evolution of the host galaxy must be quantified. 
This can be achieved by either detecting genuinely non-AGN 
galaxies, previously unidentified, in new submm blank field
surveys, or by making pointed submm observations of  samples 
covering a broad range of AGN luminosity and redshift. 
We describe the latter of these alternatives below.

\section{Submm observations of radio samples covering the P-z plane}

We have chosen to address some of the questions of evolution and formation of
massive galaxies by observing complete sub-samples of radio galaxies that
span a factor of $\sim 1000$ in radio luminosity (log\,P$_{408 MHz} 
\sim 25.0 - 28.0$\,WHz$^{-1}$sr$^{-1}$) and range in redshift between
$1 < z < 5$. The samples drawn from 3C, 6C, 7C and LBDS surveys are
shown in fig.\,3, and taken together eliminate the tight correlation between 
radio luminosity and redshift observed in any single survey by covering the
entire P-z plane. It is well known that powerful low-z radio sources reside
in elliptical galaxies and, assuming the same to be true at high-z, we
can therefore study the gas mass fraction and level of star formation in
massive host galaxies as a function of AGN luminosity at a particular
redshift, and follow the evolution of these quantities over a period of
$\sim 4$Gyr ({\it i.e.} from $z \sim 5 \rightarrow 1$). 
We will report shortly on the first JCMT observations of  
these samples at 850$\mu m$ and 450$\mu m$ using the new bolometer array SCUBA.

\begin{figure}
\centerline{
\psfig{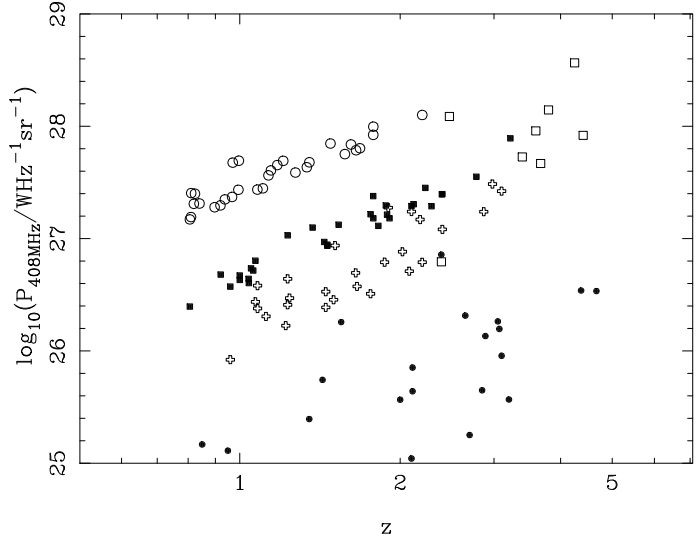}
}
\caption{\label{fig} 
The radio-luminosity:redshift ($P - z$) plane of high-z galaxies selected from
various radio surveys (3C - open circles; 6C - filled squares; 7C -
crosses; LBDS - filled circles) to be observed at submm
wavelengths.
Using these samples we can quantify the contribution of an AGN to the rest-frame FIR   
luminosity, and trace the evolution of gas mass and star formation rate
as function of redshift and radio luminosity.
} 
\end{figure}
 
\section{Acknowledgements}
It is with great pleasure that we thank our collaborators 
Steve Eales and Steve Rawlings for their contributions to this project.

\section{References}
Cimatti A., Freudling W., 1995, A\&A, 300, 366 \\
Chini R., Kr\"{u}gel E., 1994, A\&A, 288, L33 \\
Chini R., Kreysa E., Biermann P.L, 1989, A\&A, 219,87 \\ 
Dunlop J.S., Hughes D.H., Rawlings S. Eales S.A., Ward M.J., 1994, 

Nature, 370, 347 \\
de Kool M., Begelman M.C., 1989, Nature, {\bf 338}, 484 \\
Downes D., Solomom P.M., Sanders D.B., Evans A.S., 1996, A\&A
,{\bf 313}, 91 \\
Hughes D.H., Robson E.I., Dunlop J.S., Gear W.K., 1993, MNRAS, {\bf
263}, 607
Hughes D.H. 1996, in {\em Cold Gas at High Redshift}, ASSL vol 206.,
p.311, 

eds. M.N.Bremer {\it et al.}, Kluwer\\
Hughes D.H., Dunlop J.S., Rawlings S., 1997, MNRAS, 289, 766\\
Isaak K., McMahon R., Hills R., Withington, S., 1994, 269, L28 \\
Ivison R.J., 1995, MNRAS, 275, L33 \\
McMahon R.G. {\it et al.} 1994, MNRAS, 267, L9 \\
Omont A. {\it et al.} 1996, A\&A, 315, 1 \\ 
Schlickeiser R., Biermann P.L., Crusius-W\"{a}tzel A., 1991, A\&A, 247,
283 \\ 
Taylor G.L., Dunlop J.S., Hughes D.H., Robson E.I., 1996, MNRAS, 283,
930 \\
\end{document}